\documentclass[aip,reprint,showpacs,prb,amssymb,amsmath,twocolumn,floatfix]{revtex4}
\usepackage{graphicx}% Include figure files
\usepackage{dcolumn}% Align table columns on decimal point
\usepackage{bm}% bold math
\usepackage{color}
\begin{document}

\title{Transport Properties of Cubic Crystalline Ge$_{2}$Sb$_{2}$Te$_{5}$: A Potential Low-temperature Thermoelectric Material}% Force line breaks with \\

\author{Jifeng Sun$^{1,2,3}$}
\author{Saikat Mukhopadhyay$^{1}$}
\author{Alaska Subedi$^{4}$}
\author{Theo Siegrist$^{2,3}$}
\author{David J. Singh$^{1}$}
%\email{huipan@umac.mo}
 \affiliation{$^1$Materials Science and Technology Division, Oak Ridge National Laboratory, Oak Ridge, TN 37831, USA}
 \affiliation{$^2$Florida Agricultural $\&$ Mechanical University-Florida State University, College of Engineering, Department of Chemical $\&$ Biomedical Engineering, 2525 Pottsdamer St., Tallahassee, FL 32310, USA}
  \affiliation{$^3$National High Magnetic Field Laboratory, 1800 E. Paul Dirac Dr., Tallahassee, FL 32310, USA}
  \affiliation{$^4$Max Planck Institute for the Structure and Dynamics of Matter, Hamburg, Germany}

\date{\today}% It is always \today, today,

\begin{abstract}
Ge$_{2}$Sb$_{2}$Te$_{5}$ (GST) has been widely used as a popular phase change material. In this study, we show that it exhibits high Seebeck coefficients 200 - 300 $\mu$V/K in its cubic crystalline phase ($\it{c}$-GST) at remarkably high $\it{p}$-type doping levels of $\sim$ 1$\times$10$^{19}$ - 6$\times$10$^{19}$ cm$^{-3}$ at room temperature. More importantly, at low temperature (T = 200 K), the Seebeck coefficient was found to exceed 200 $\mu$V/K for a doping range 1$\times$10$^{19}$ - 3.5$\times$10$^{19}$ cm$^{-3}$. Given that the lattice thermal conductivity in this phase has already been measured to be extremely low ($\sim$ 0.7 W/m-K at 300 K),\citep{r51} our results suggest the possibility of using $\it{c}$-GST as a low-temperature thermoelectric material.

\end{abstract}

\pacs{Computational Physics, Condensed Matter Physics}% PACS, the Physics and Astronomy
%\keywords{Suggested keywords}%Use show keys class option if keyword display desired
\maketitle

\section{\label{sec:level1}INTRODUCTION}

Low-temperature thermoelectric (TE) materials are important for refrigeration applications in electronics, infrared detectors, computers and other areas. \citep{r49} However, in contrast to high temperature TE materials, there is relatively little progress in the discovery of high performance low temperature TE materials. The most widely used materials at ambient temperature and below are (Bi,Sb)$_{2}$Te$_{3}$ derived compounds. \cite{r7,r8} These have figures of merit, ZT, of roughly unity at 300 K with lower ZT values at lower temperatures. Here we show that cubic GST phase change material has potential to become an excellent thermoelectric material in the important temperature range from 200 K - 300 K. Specifically, we find using Boltzmann transport calculations that disordered cubic $\it{p}$-type GST will have high thermopowers in the range 200 $\mu$V/K - 300 $\mu$V/K at remarkably high doping levels consistent with good conductivity. GST already is known to have a very low thermal conductivity. \cite{r51}  

Pseudobinary alloys of (GeTe)$_{m}$-(Sb$_{2}$Te$_{3}$)$_{n}$ have been used in data storage for more than a decade because of their fast phase switching between metastable crystalline (cubic) and amorphous phases.\citep{r9,r10,r11} The above room temperature thermoelectric properties of (GeTe)$_{m}$-(Sb$_{2}$Te$_{3}$)$_{n}$ were previously studied experimentally.\citep{r12,r13,r14,r15,r16} In particular, for m = 2 and n = 1; ordered Ge$_{2}$Sb$_{2}$Te$_{5}$ was reported to  show interesting thermoelectric properties, even at room temperature. \citep{r13,r15,r16} 

Disorder in the crystal structure can be beneficial for thermoelectric properties \citep{r44,r45} and hexagonal GST has been already reported to have reasonably good thermoelectric properties. \citep{r12,r43} The disorder in phase change materials was discussing in detail by Zhang and co-workers. \citep{r58} We therefore investigate the thermoelectric properties of cubic disordered GST. Previous measurements of thermoelectric properties in $\it{c}$-GST were restricted to thin-film samples where the thermoelectric properties were found to vary significantly depending on the thickness of the sample.\citep{r15,r34} The thermoelectric properties of $\it{c}$-GST in the bulk form have remained unknown. We note that Lee and co-workers have discussed resonant bonding in the context of thermoelectric performance \citep{r55} and that resonant bonding has been identified in phase change materials. \citep{r56,r57}

\section{\label{sec:level2}THEORETICAL APPROACH AND COMPUTATIONAL DETAILS}
The ab-initio calculations presented here were carried out using the linearized augmented plane-wave (LAPW) method \citep{r20} as implemented in the WIEN2K code.\citep{r21} The structures were relaxed using the PBE-GGA functional then a semilocal functional of Tran and Blaha (TB-mBJ) \citep{r22} was employed to treat the exchange-correlation potential in the calculations of electronic, optical and transport properties. This functional form was shown to better describe the band gaps of a variety of semiconductors and insulators.\citep{r23,r24}
 
The cubic phase of GST occurs in a rock-salt structure with Te occupying one site and Ge, Sb and vacancies randomly occupying the other site. We generated special quasirandom structures (SQS) using the prescription of Zunger \textit{et al.}\cite{r39} for our first principles study.  The SQSs were generated such that all the pair correlation functions (PCF) up to the next nearest neighbors are identical to the average PCF of an infinite random structure. We used the \texttt{mcsqs} code as implemented in ATAT software package \citep{r48} to generate the SQSs using a Monte Carlo algorithm. We considered two SQS structures with 27 atoms (6 Ge, 6 Sb, and 15 Te atoms) and 45 atoms  (10 Ge, 10 Sb, and 25 Te atoms)  in the unit cell and compared their electronic and transport properties. Despite the relatively poor quality of PCFs for the case with 27-atomic unit cell, we find that the overall calculated properties for these structures are very similar (a detailed comparison of electronic, optical and transport properties in these structures can be found in Ref. \citep{r54}). This may indicate that the inherent properties in this phase is independent of its PCFs. Thus this 45-atomic unit cell is converged with respect to size to calculate properties of $\it{c}$-GST.
 
The self-consistent calculations were performed with a 4$\times$4$\times$4 k-point grid of 36 k points in the irreducible Brillouin zone (IBZ) whereas electronic and optical properties were calculated using a 8$\times$8$\times$8 k-mesh. For the calculation of the thermoelectric properties, we employed a denser k-point mesh of 14$\times$14$\times$14. All the calculations were duly tested for convergence with a variety of  k-point grids. The thermopower was analysed using the Boltztrap code \citep{r25} employing Boltzmann transport theory within the constant scattering time approximation (CSTA). This CSTA approach has been shown to be successful in calculating the Seebeck coefficient in a variety of thermoelectric materials.\citep{r26,r35,r36,r37,r38} The substance of CSTA is to assume that the energy dependence of the scattering rate is small compared with the energy dependence of the electronic structure. A detailed description of this approach can be found elsewhere.\citep{r27,r35} We used well converged basis sets with LAPW basis size corresponding  to R$_{MT}$K$_{max}$=9.0, where R$_{MT}$ is the smallest Muffin Tin radius and K$_{max}$ is the plane-wave cutoff parameter. The LAPW sphere radius of 2.47 Bohr was used for Ge whereas 2.5 Bohr was used for Sb and Te. Spin-orbit coupling  was included for all the properties reported.

\section{\label{sec:level3}RESULTS AND DISCUSSIONS}

\subsection{\label{sec:1}Electronic structure and optical properties}

\begin{figure}
\includegraphics*[height=6cm,keepaspectratio]{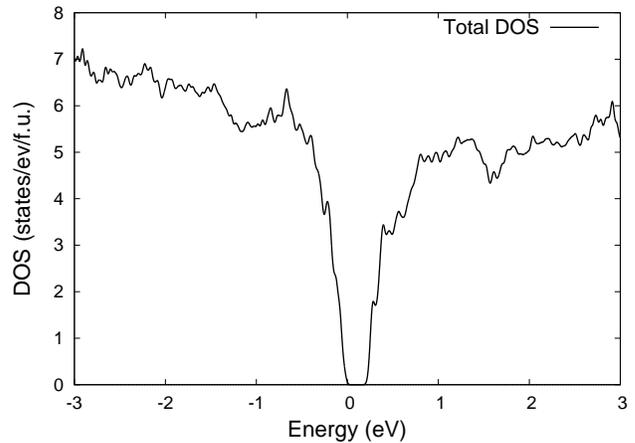}% Here is how to import EPS art
\caption{\label{fig1}Calculated total density of states for $\it{c}$-GST. The Fermi level is set to zero.}
\end{figure}

The total density of states (DOS) of $\it{c}$-GST near the Fermi energy (E$_{F}$) is shown in Fig. \ref{fig1}.  We find a clear semiconducting behavior with a band gap of 0.17 eV. This is in good agreement with a previous theoretical investigation\citep{r30} where an indirect band gap of 0.1 eV was noted for $\it{c}$-GST although they used different exchange-correlation potential (local density approximation without spin-orbit coupling) and different method (shifting of the hexagonal structure) to generate the cubic structure. Another previous study considered the cubic GST as a narrow-gap degenerate semiconductor, with E$_{F}$ reside inside the valence band or a defect band.\citep{r29} The experimentally reported estimated optical gap is $\sim$ 0.5 eV \citep{r29,r33} which is larger than what we obtain.

\begin{figure}
\includegraphics*[height=6cm,keepaspectratio]{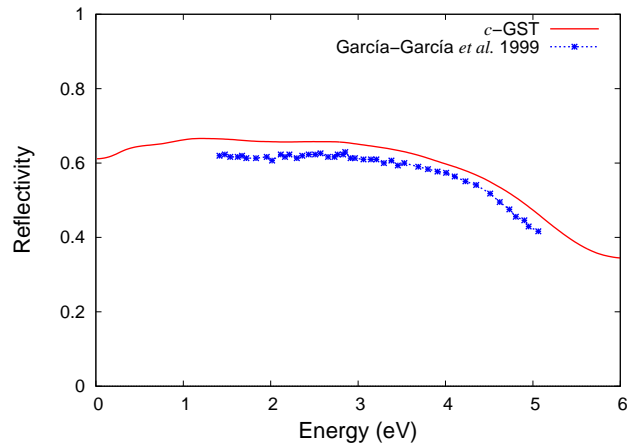}% Here is how to import EPS art
\caption{\label{fig2}Calculated reflectivity of $\it{c}$-GST. Previous experimental result from Ref. \citep{r32} is also shown for a better comparison.}
\end{figure}

\begin{figure}
\includegraphics*[height=6cm,keepaspectratio]{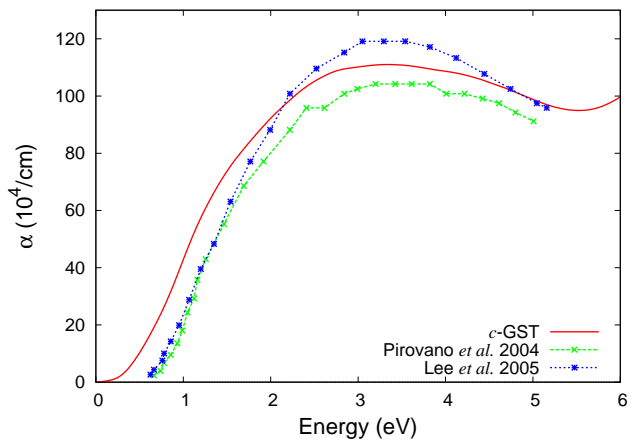}% Here is how to import EPS art
\caption{\label{fig3} Calculated optical absorption coefficient along with experimental data from  previous measurements (see Refs. \citep{r29,r33})}
\end{figure}

GST has applications in rewritable DVD and Blu-ray technologies where a large contrast between the optical properties of the cubic and amorphous phases is exploited.\citep{r9,r11} In Fig. \ref{fig2} and \ref{fig3}, we present the averaged reflectivity and optical absorption coefficients for $\it{c}$-GST, respectively.  It is evident from Fig. \ref{fig2} that $\it{c}$-GST displays a high reflectivity ranging from 0.65 to 0.68 within the energy up to 3.5 eV which covers the energy range of DVD (1.91 eV) and blu-ray (3.06 eV)\citep{r9} technology . This is consistent with the measured optical reflectivity by Garcia-Garcia ${et}$ ${al.}$'s \citep{r32} [Fig. \ref{fig2}], though they used thin films of  $\it{c}$-GST in their experiments. Fig. \ref{fig3}. depicts the calculated optical absorption coefficients together with that from previous measurements. Our calculated absorption coefficient compares well in magnitude with that from Lee ${et}$ ${al.}$ \citep{r29} and Pirovano ${et}$ ${al.}$ \citep{r33} for the higher energy regime but disagrees at lower energy reflecting the band gap difference. One possible explanation is that the structure is more complex than the random structure that we assumed, or that larger length scale effects such as local near ferroelectric polarizations play a role in the samples. Additionally, the dimension of the crystals may also play a role since thin films of GST were used in both of the experiments instead of the ideal bulk that we are considering.

\subsection{\label{sec:2}Thermoelectric properties}

\begin{figure}
\includegraphics*[height=6cm,keepaspectratio]{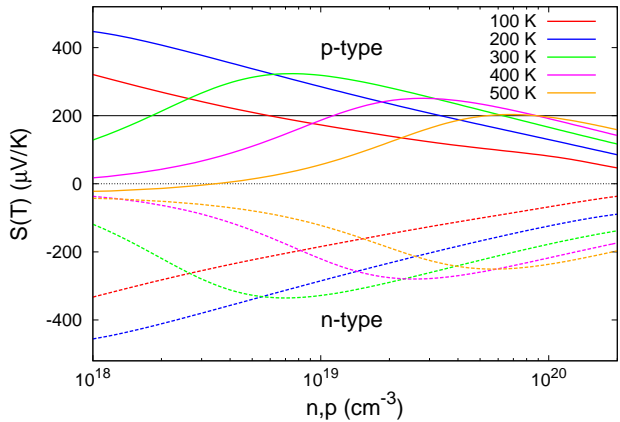}% Here is how to import EPS art
\caption{\label{fig4} Calculated thermopowers of $\it{c}$-GST as a function of doping concentration both for $\it{p}$-type and $\it{n}$-type at various temperatures. Solid horizontal line at 200 $\mu$V/K represents the limitation of Seebeck coefficients for good TE materials.}
\end{figure}

In Fig. \ref{fig4}, we have plotted the calculated thermopower (S(T)) of $\it{c}$-GST at various temperatures ranging from 100 K up to 500 K, for both electron-doped ($\it{n}$-type) and hole-doped ($\it{p}$-type) cases. It has been reported that it might be possible to dope GST $\it{n}$-type\citep{r52,r53}, however, keeping in mind that most of the experiments deal with $\it{p}$-type GST, we focus on the thermoelectric properties of $\it{p}$-type $\it{c}$-GST. In fact, there is a rather good electron-hole symmetry in the S(T), so $\it{n}$-type if realized would be similar to $\it{p}$-type. We give the direction average  of S(T), i.e. S=$\frac{S_{xx}+S_{yy}+S_{zz}}{3}$, to eliminate artificial anisotropy due to the unit supercell.

We find $\it{c}$-GST exhibits a Seebeck coefficients 200 - 323 $\mu$V/K when doped $\it{p}$-type for doping concentration 1.7$\times$10$^{18}$ - 6.1$\times$10$^{19}$ cm$^{-3}$ at room temperature. More importantly, at low temperature (T = 200 K), the Seebeck coefficient exceeds 200 $\mu$V/K for a doping range of 1$\times$10$^{19}$ - 3.5$\times$10$^{19}$ cm$^{-3}$. A strong bipolar effect was noted for T = 300 K and higher temperatures, however, it was not present at lower temperatures. This is connected with the band gap, which as noted may be too low in the present calculations. In order to assess the possibility of $\it{c}$-GST  as a low temperature TE material, it is necessary to compare the calculated Seebeck coefficient with that in popular low-temperature TE materials. For example, Poudel and co-workers\citep{r7} reported a bulk nano-composite Bi$_{x}$Sb$_{2-x}$Te$_{3}$ with S(T) of 185 $\mu$V/K at 300 K. Chung ${et}$ ${al.}$ \citep{r8} discovered CsBi$_{4}$Te$_{6}$ with a thermopower of 105 $\mu$V/K at 300 K. Importantly, the high thermopowers are at high carrier concentrations where reasonable electric conductivity may be anticipated. 

\begin{figure}
\includegraphics*[height=6cm,keepaspectratio]{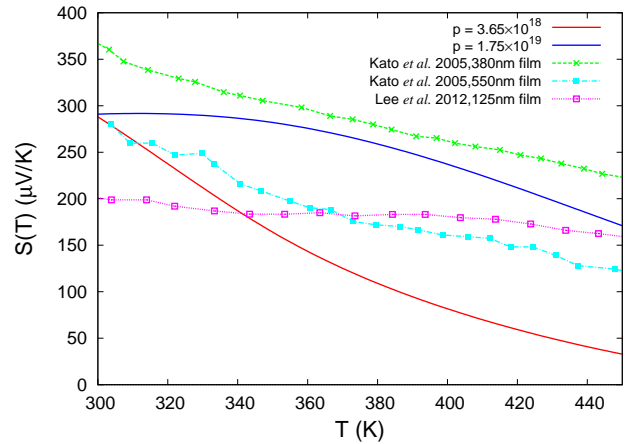}% Here is how to import EPS art
\caption{\label{fig5}Temperature dependence of the thermopower of $\it{c}$-GST with two $\it{p}$-type doping concentrations and experimental results from previous work. \cite{r15,r34}}
\end{figure}

We shall now focus on comparing our calculated results with previous experiments on GST thin films. Kato and Tanaka \citep{r34} investigated thermoelectric properties of Ge$_{2}$Sb$_{2}$Te$_{5}$ thin films with sample dimensions ranging from 50 nm - 600 nm for a temperature range 25 $^{\circ}\mathrm{C}$ to 200 $^{\circ}\mathrm{C}$. For their $\it{c}$-GST samples, they found a pretty high values of S(T) from 280 $\mu$V/K to 380 $\mu$V/K. Lee ${et}$ ${al.}$ \citep{r15} also measured the temperature dependence of S(T) with 25 nm and 125 nm Ge$_{2}$Sb$_{2}$Te$_{5}$ thin films from 20 $^{\circ}\mathrm{C}$ to 300 $^{\circ}\mathrm{C}$ and observed S(T) as high as 205 $\mu$V/K and 175 $\mu$V/K for 125 nm and 25 nm thick films, respectively. Both of the authors found a decreasing S(T) with increasing temperatures. In order to further compare our calculations with these measurements, we matched the calculated Seebeck coefficients at 300 K with that from Kato's data for the 550 nm thick thin film and calculated the corresponding doping levels in $\it{c}$-GST. For these particular doping levels, we then plotted the variation of S(T) with temperature together with the respective experimental data, as shown in Fig. \ref{fig5}. As seen, a similar decreasing trend of S(T) with respect to temperatures can be found in the calculation for both doping concentrations. However, the agreement of the decreasing feature is more obvious in the low doping level case which falls into the bipolar region and the deviation may stem from the underestimation of the band gap. This may also suggest the fact that the experimental samples are in the low carrier concentrations range where bipolar effect is strong and detrimental to the thermoelectric properties. Therefore a better thermoelectric performance might be feasible in higher doping concentration samples.

% and may cause difficulties in predicting the real behavior of the thermopower comparing to experiments. On the other hand, for the higher doping level, the thermopower starts to decrease at T $>$ 340 K which may indicate the fact that this doping concentration, though not in the bipolar region, is close to the maximum value and cause a analogous decreasing character at higher temperatures.

%In our calculations, we find a decreasing feature of S(T) in low doping level cases which fall into the bipolar regions, but with bigger decreasing slopes and this large deviation may due to the small band gap. Moreover, the bipolar effect is detrimental to the thermoelectric properties and may cause difficulties in predicting the real behavior of the thermopower comparing to experiments. 

%We find an acceptable agreement between the calculated and measured S(T) in the high doping levels for both cases.  The small deviation may stem from many aspects such as the quality of the crystal and processing methods, as well as the sample dimensions. However, the continuous decrease in S(T) that was noted in both the reports was found to occur at higher temperature ($>$ 450 K) in our calculations. On the other extreme, at lower doping concentrations,  the calculated S(T)-T could not explain what was observed in the measurement. This may be due to the fact that this carrier concentration falls under the range where we find  a strong bipolar effect in our calculated S(T). 

\section{\label{sec:level4}SUMMARY$\&$CONCLUSIONS}

To summarize, we performed first-principles calculations to investigate electronic, optical and thermoelectric properties of $\it{c}$-GST with fully random cation site occupancy generated using the SQS scheme. Our results based on Boltzmann transport calculations predict substantially high thermopower at high carrier concentration ($\sim$ 300 $\mu$V/K at 7$\times10^{18}$ cm$^{-3}$) at room temperature. We also find high thermopower at high carrier concentrations going down to 200 K and below suggesting that with heavy hole doping $\it{c}$-GST may be an excellent low-temperature thermoelectric. These results suggest future experiments to better understand the low-temperature thermoelectric properties of $\it{c}$-GST. Specifically, it would be of interest to study the temperature range of 200 K - 300 K with high doping concentrations ranging from 1$\times10^{19}$ cm$^{-3}$ up to 6$\times10^{19}$ cm$^{-3}$.  

\begin{acknowledgments}
This work was supported by the Department of Energy through the S3TEC energy frontier research center. JS acknowledges a graduate student fellowship, funded by the Department of Energy, Basic Energy Science, Materials Sciences and Engineering Division, through the ORNL GO! program. A portion of this work was performed at high-performance computing center at the National High Magnetic Field Laboratory.
\end{acknowledgments}

\bibliography{c-GST}

\end{document}